\documentclass[english,aps,prd,showpacs,twocolumn]{revtex4-1}
\usepackage[T1]{fontenc}
\usepackage[latin9]{inputenc}
\usepackage{xcolor}
\usepackage{pdfcolmk}
\usepackage{textcomp}
\usepackage{amsmath}
\usepackage{amssymb}
\usepackage{graphicx}
\PassOptionsToPackage{normalem}{ulem}
\usepackage{ulem}

\makeatletter

\providecommand{\tabularnewline}{\\}
\providecolor{lyxadded}{rgb}{0,0,1}
\providecolor{lyxdeleted}{rgb}{1,0,0}
\usepackage{slashed}

\makeatother

\usepackage{babel}
\begin{document}

\title{Topological responses from chiral anomaly in multi-Weyl semimetals}

\author{Ze-Min Huang}

\affiliation{Department of Physics, The University of Hong Kong, Pokfulam Road,
Hong Kong, China}

\author{Jianhui Zhou}

\affiliation{Department of Physics, The University of Hong Kong, Pokfulam Road,
Hong Kong, China}

\author{Shun-Qing Shen}

\affiliation{Department of Physics, The University of Hong Kong, Pokfulam Road,
Hong Kong, China}

\date{\today }
\begin{abstract}
Multi-Weyl semimetals are a kind of topological phase of matter with
discrete Weyl nodes characterized by multiple monopole charges, in
which the chiral anomaly, the anomalous nonconservation of an axial
current, occurs in the presence of electric and magnetic fields. Electronic
transport properties related to the chiral anomaly in the presence
of both electromagnetic fields and axial electromagnetic fields in
multi-Weyl semimetals are systematically studied. It has been found
that the anomalous Hall conductivity has a modification linear in
the axial vector potential from inhomogeneous strains. The axial electric
field leads to an axial Hall current that is proportional to the distance
of Weyl nodes in momentum space. This axial current may generate chirality
accumulation of Weyl fermions through delicately engineering the axial
electromagnetic fields even in the absence of external electromagnetic
fields. Therefore, this work provides a nonmagnetic mechanism of generation
of chirality accumulation in Weyl semimetals and might shed new light
on the application of Weyl semimetals in the emerging field of valleytronics.
\end{abstract}

\maketitle

\section{introduction}

Weyl semimetals are materials with a finite number of band touching
points, Weyl nodes, in the three-dimensional Brillouin zone \cite{hosur2013crp,WengHM2016jpcm,burkov2017arcmp,Armitage2017rmp}.
These Weyl nodes can be viewed as magnetic monopoles in momentum space
\cite{Xiao10RMP,volovik2003oxford}, which can lead to various anomalous
transport properties, including static and dynamical chiral magnetic
effects \cite{vilenkin1980prd,fukushima2008prd,Grushin2012PRD,zyuzin2012prb,zhou2013cpl,vazifeh2013prl,Goswami2013PRBaxion,landsteiner2014PRB,chang2015prb,MaPesin2015PRB,zhongPRL2016,Alavirad2016PRB,Baireuther2016nkp,MaPesin2017PRL},
anomalous Hall effect (AHE) \cite{shindou2001prl,yang2011prb}, chirality-dependent
Hall effect \cite{YangSY2015PRL,JiangQD2015PRL}, nonlocal transport
signature \cite{ParameswaranPRX2014} and anomalous magnetoresistance
\cite{nielsen1983plb,Son2013prb,KimHJ2013PRL,burkov2014prl,Xiong2015Science,Huang2015PRX,li2015NC,LiH2016NC,Li2016NP,Zhang2016nc}.
However, most of these studies focused on single-Weyl semimetals,
whose energy dispersions are linear in wave vectors and monopole charge
equals $\pm1$. Recently, there are proposals of multi-Weyl fermions
\cite{xu2011prl,fang2012prl,huang2016pnas}, with multiple monopole
charges and nonlinear dispersion relation. For example, first principle
calculations suggest that a pair of double-Weyl nodes exist in $\mathrm{HgCr_{2}Se_{4}}$
\cite{xu2011prl} and $\mathrm{SrSi_{2}}$ \cite{huang2016pnas}.
It has been shown that the nonlinear dispersion and double monopole
charges of double-Weyl semimetals have led to unconventional correlation
effects \cite{jian2015prb,lai2015prb,Ahn2016sr,roy2017prb}, magnetotransport
\cite{chang2015prb,daiXin2016prb,lixiao2016prb,Park2017PRB} and thermoelectric
transport \cite{ChenQi2016PRB}. In this work, our prime aim is to
systematically investigate topological transport properties of multi-Weyl
fermions associated with the chiral anomaly.

The chiral anomaly in single-Weyl semimetals had been derived in several
different ways \cite{nielsen1983plb,bertlmann2000book,fujikawa2004oxford,Son2013PRD},
but derivation of the chiral anomaly in multi-Weyl semimetals is still
absent. In the intuitive and physical derivation given by Nielsen
and Ninomiya \cite{nielsen1983plb}, helicity plays an important role.
Namely, for massless fermions in a homogeneous magnetic field $\mathbf{B}$,
the spins are preferentially aligned along $\mathbf{B}$. Then the
left-handed and the right-handed fermions are accelerated under an
electric field $\mathbf{E}$, resulting in Adler-Bell-Jackiw anomaly
or chiral anomaly \cite{bertlmann2000book}. When it comes to multi-Weyl
semimetals, the Lorentz invariance is broken such that helicity is
not a well-defined quantity. Hence, the specific forms of the chiral
anomaly and Jacobian under a chiral transformation are still absent,
which call for a detailed derivation. Since the chiral anomaly closely
relates to topological responses \cite{zyuzin2012prb,hosur2013crp},
the chiral anomaly shall assist us to construct the corresponding
effective action and then to investigate the relevant topological
responses.

In addition, axial magnetic fields, as large as 300 tesla, can be
simulated by elastic deformations of lattice and couple to Dirac fermions
in graphene \cite{levy2010science,jackiw2007prl,vozmediano2010pr}.
For single-Weyl semimetals, three-dimensional counterparts of graphene,
it has been demonstrated that axial magnetic fields emerge from strain
fields, nonuniform magnetizations or topological defects \cite{liu2013prb,cortijo2015prl,liu2017prb}.
Theoretical investigations show that the axial gauge fields can induce
anomalous topological properties in single-Weyl semimetals, including
chiral pseudomagnetic effect \cite{zhou2013cpl,Sumiyoshi2016PRL,grushin2016prx,pikulin2016prx},
plasmon-magnon coupling \cite{liu2013prb}, phonon Hall viscosity
\cite{cortijo2015prl,shapourian2015prb}, emergent gravity \cite{Zubkov2015ap,chernodub2017prb},
and chiral magnetic plasmons \cite{Gorbar2017prl}. Therefore, one
would naturally wonder whether the interplay between electromagnetic
fields and axial gauge fields occurs in multi-Weyl semimetals.

In this paper, we apply the quantum-field-theory approach to explore
the topological responses of multi-Weyl semimetals to the electromagnetic
fields as well as the axial electromagnetic fields. We find that the
strain fields make a significant contribution linear in the axial
vector potential $\mathbf{A}^{5}$ to the anomalous Hall conductivity.
It has been shown that both an axial magnetic field and an axial electric
field lead to axial currents proportional to the chiral chemical potential
$b_{0}$ and the distance of Weyl nodes in momentum space $\mathbf{b}$,
respectively. Consequently, the chirality accumulation of Weyl fermions
can be achieved at the surfaces of a sample completely through engineering
the axial fields. In addition, the realization and detection of the
strain-induced AHE, the axial currents and the resulting chirality
accumulation in double-Weyl semimetals are discussed.

The rest of this paper is organized as follows. In Sec. \ref{sec:Model},
we introduce the effective Hamiltonian and the Lagrangian density
for multi-Weyl semimetals and discuss the topological invariant. In
Sec. \ref{sec:Chiral-anomaly}, we derive the chiral anomaly equations
for multi-Weyl semimetals in the presence of both the electromagnetic
fields and the axial gauge fields. In Sec. \ref{sec:top-resp}, the
topological responses due to the chiral anomaly are present. The realizations
and detections of the strain-induced AHE, electric/axial currents
and chirality accumulation in double-Weyl semimetals are proposed.
In Sec. \ref{sec:conclusions}, the main results of this paper are
summarized. Finally, in Appendixes, we give a detailed calculations
of the effective action and the chiral anomaly equations.

\begin{table}
\begin{tabular}{|c|c|}
\hline
$n$ & DOS\tabularnewline
\hline
\hline
1 & $\frac{w^{2}}{2\pi^{2}v^{3}}E^{2}$\tabularnewline
\hline
2 & $\frac{\sqrt{2}w}{8\pi v^{2}}\left|E\right|$\tabularnewline
\hline
3 & $\left[\frac{w^{2/3}\Gamma(1/3)^{3}}{2^{8/3}\sqrt{3}\pi^{3}v^{5/3}}\right]|E|^{2/3}$\tabularnewline
\hline
\end{tabular}\caption{Monopole charge $n$ and density of states at zero temperature for
Weyl semimetals with $n=1,2,3$. $\Gamma\left(x\right)$ is the gamma
function.}
\label{DOSAn}
\end{table}

\section{model for multi-Weyl semimetals \label{sec:Model}}

We start with the following Hamiltonian for multi-Weyl semimetals
containing a pair of Weyl nodes with multiple monopole charges \cite{volovik1988sjetp,fang2012prl,lixiao2016prb}
\[
H=\left(\begin{array}{cc}
H_{+} & 0\\
0 & H_{-}
\end{array}\right),
\]
with
\begin{eqnarray}
H_{s} & = & sv\Big[(\mathbf{p}+s\mathbf{b})_{3}\sigma^{3}\nonumber \\
 &  & +w^{-1}(\mathbf{p}+s\mathbf{b})_{+}^{n}\sigma^{-}+w^{-1}(\mathbf{p}+s\mathbf{b})_{-}^{n}\sigma^{+}\Big].\label{eq:hamiltonian}
\end{eqnarray}
where $v$ is the effective velocity, $w$ is a material-dependent
parameter, $\mathbf{p}$ denotes for momentum, $p_{\pm}=\left(p_{1}\pm ip_{2}\right)/\sqrt{2}$,
and $\mathbf{b}$ characterizes the distance between Weyl nodes with
opposite chirality in momentum space. $s=\pm1$ are the chiralities
of Weyl nodes. The Weyl node $s$ locates at $-s{\bf b}$ in momentum
space. $\sigma^{i}$ are the Pauli matrices $\left(i=1,\thinspace2,\thinspace3\right)$,
$\sigma^{\pm}=(\sigma^{1}\pm i\sigma^{2})/\sqrt{2}$ and $n$ is a
positive integer. It has been pointed out that both double-($n=2$)
and triple-($n=3$) Weyl semimetals are protected by $C_{4}$ and
$C_{6}$ symmetry, respectively. But other more higher order band-crossing
points are not protected by $n$-fold rotational symmetry \cite{fang2012prl}.
Thus, the possible values of $n$ for multi-Weyl semimetals are $n=2,3$,
which might be detected by quantum transport measurement \cite{daiXin2016prb}.
Their corresponding density of states (DOS) is present in Table \ref{DOSAn}.
One finds that, compared with single-Weyl semimetals, the DOS of either
double- or triple-Weyl semimetals possesses greatly different dependence
of energy. To be specific, the DOS obeys the law $\left|E\right|^{2/n}$
up to a material-dependent constant. In addition, the DOS for Weyl
semimetals with $n=1,2,3$ vanishes identically at Weyl nodes.

For each Weyl node, the winding number can be defined by \cite{volovik2003oxford}
\begin{equation}
N_{s}=-\frac{1}{3!(2\pi i)^{2}}\int_{S^{3}}\text{tr}(G^{\tau}dG^{\tau-1})^{3},\label{eq:windernumber}
\end{equation}
where $G^{\tau}$ is the imaginary-time Green's function and $S^{3}$
means integrating over a three-dimensional sphere in frequency-momentum
space enclose this Weyl node. $\text{tr}$ acts on the degree of freedom
$\sigma^{i}$. For our model in Eq. $\left(\ref{eq:hamiltonian}\right)$,
one finds
\begin{equation}
N_{s}=sn.\label{Ns}
\end{equation}
Although the winding number $N_{s}$ defined here is equivalent to
the Chern number \cite{TKNN1982PRL}, to make a closer connection
to the effective action below, we shall adopt the winding number rather
than the Chern number throughout this paper. The sum of $N_{s}$ $\left(N=\Sigma_{s}N_{s}\right)$
can also be obtained from integrating over the whole area $p_{0,1,2}\in(-\infty,\infty)$
with fixed $p_{3}$
\begin{equation}
N(p_{3})=n\left[\theta\left(p_{3}+b_{3}\right)-\theta\left(p_{3}-b_{3}\right)\right],\label{eq:Chern}
\end{equation}
where $\theta(x)$ is the Heaviside step function: $\theta(x)=1$
for $x\geqslant0$ and otherwise vanishes. Eq. $\left(\ref{eq:Chern}\right)$
shall manifest itself in the effective action. From Eq. $\left(\ref{eq:Chern}\right)$,
one can determine the locations of Weyl nodes: $\frac{dN}{dp_{3}}=\pm n\delta(p_{3}\pm b_{3}).$

As mentioned above, due to the explicit violation of Lorentz symmetry,
the helicity is not well-defined and the chiral condition is invalid
as well: $\gamma^{5}\Psi_{\pm}=\pm\Psi_{\pm}$. For single-Weyl nodes,
the chirality is equivalent to the sign of the winding number. Analogically,
we adopt the latter as a generalized definition for the chirality.
With this respect, the $\gamma^{5}$ matrix can be defined by setting
$s$ as the eigenvalue and $\Psi_{\pm}$ as the eigenfunctions.
\begin{widetext}
For later convenience, we write the corresponding Lagrangian density
as
\begin{eqnarray}
\mathcal{L} & = & \bar{\Psi}\{\gamma^{0}(p_{0}-eA_{0}-eA_{0}^{5}\gamma^{5}+m_{0}+b_{0}\gamma^{5})+\gamma^{3}(p_{3}-eA_{3}-eA_{3}^{5}\gamma^{5}+m_{3}+b_{3}\gamma^{5})\nonumber \\
 &  & +\frac{1}{w}\left[\gamma^{+}\left(p-eA-eA^{5}\gamma^{5}+m+b\gamma^{5}\right)_{+}^{n}+\gamma^{-}\left(p-eA-eA^{5}\gamma^{5}+m+b\gamma^{5}\right)_{-}^{n}\right]\}\Psi,\label{eq:lagrangian}
\end{eqnarray}
where $\mu,\thinspace\nu=0,\thinspace1,\thinspace2,\thinspace3$,
$v$ has been absorbed into $p_{\mu}=\left(p_{0},-\mathbf{p}\right)$,
$b_{0}$ is the chiral chemical potential in a steady state not the energy difference between two Weyl nodes~\cite{fukushima2008prd,Zhou2015PRB},
$m_{0}$ is the averaged chemical potential and $-\mathbf{m}$ is
the center of momentum of Weyl nodes. $\mu_{s}=m_{0}+sb_{0}$ is the
chirality-dependent chemical potential. $\gamma^{\mu}$ is the gamma
matrix, satisfying $\left\{ \gamma^{\mu},\thinspace\thinspace\gamma^{\nu}\right\} =2g^{\mu\nu}$
with $g^{\mu\nu}=\mathrm{diag}\left\{ 1,-1,-1,-1\right\} $, $\gamma^{i}=\sigma^{i}\otimes\left(-i\tau^{2}\right)$,
$\gamma^{0}=\sigma^{0}\otimes\tau^{1}$, $\gamma^{\pm}=\left(\gamma^{1}\mp i\gamma^{2}\right)/\sqrt{2}$,
and $\gamma^{5}=\sigma^{0}\otimes\tau^{3}$. $\Psi$ is a Grassmann
number, $\Psi=(\Psi_{+},\ \Psi_{-})^{T}$ and $\bar{\Psi}=\Psi^{\dagger}\gamma^{0}$.
$F_{\mu\nu}=\partial_{\mu}A_{\nu}-\partial_{\nu}A_{\mu}$ is the field
strength tensor for electromagnetic fields. $A_{\mu}^{5}$ is the
axial vector potential. The field strength tensor for $A_{\mu}^{5}$
is defined as $F_{\mu\nu}^{5}=\partial_{\mu}A_{\nu}^{5}-\partial_{\nu}A_{\mu}^{5}$.
In analogy to Maxwell's equations for electromagnetic fields, it is
instructive to examine the dynamics of axial electromagnetic fields.
The Bianchi identity for $F_{\mu\nu}^{5}$ gives rise to $\nabla\cdot\mathbf{B}_{5}=0$
and $\partial_{t}\mathbf{B}_{5}=-\nabla\times\mathbf{E}_{5}$, which
are identical to the ones for $F_{\mu\nu}$. However, the sourceless
free equations of motion when $A_{0}^{5}=0$ become
\begin{eqnarray}
\nabla\cdot\mathbf{E}_{5} & = & \partial_{t}\left(\nabla\cdot\mathbf{A}^{5}\right)
\end{eqnarray}
and
\begin{eqnarray}
\partial_{t}\mathbf{E}_{5}-\nabla\times\mathbf{B}_{5} & = & \nabla\left(\nabla\cdot\mathbf{A}^{5}\right)-\nabla^{2}\mathbf{A}^{5}+\partial_{t}^{2}\mathbf{A}^{5}.
\end{eqnarray}
Since $A_{\mu}^{5}$ is observable and single valued \cite{cortijo2015prl,cortijo2016prb,pikulin2016prx},
one thus has no redundant gauge freedom to ensure $\nabla\cdot\mathbf{A}^{5}=0$.
It should been pointed out that in general, the strain fields should not only produce an axial gauge potential but also modify the local geometry metric~\cite{Zubkov2015ap,chernodub2017prb}. In this work, we mainly focus on the impact of axial gauge potentials on Weyl fermions.
Note that the 4-vector $b_{\mu}=\left(b_{0},-\mathbf{b}\right)$ can
be eliminated by performing a large chiral transformation, $\Psi\rightarrow e^{i(b_{\mu}x^{\mu})\gamma^{5}}\Psi$.
Due to Fujikawa's uncertainty principle \cite{bertlmann2000book,fujikawa2004oxford},
such a chiral transformation would give rise to a term from the path-integral
measure. However, the specific form of this Jacobian is still absent
for multi-Weyl semimetals. Note that we have neglected the conventional
action for classical electrodynamics $-F_{\mu\nu}F^{\mu\nu}/4$ in
Eq. $\left(\ref{eq:lagrangian}\right)$.

\begin{table}
\begin{tabular}{|c|c|c|}
\hline
Name & Form & References \tabularnewline
\hline
\hline
Phonon Hall viscosity & $S_{\text{eff }}=-i\frac{ne^{2}}{4\pi^{2}\hbar}\int d^{d}x\epsilon^{\mu\nu\rho\sigma}b_{\mu}A_{\nu}^{5}\partial_{\rho}A_{\sigma}^{5}$ & \cite{cortijo2015prl}\tabularnewline
\hline
charge density & $j^{0}=\frac{ne^{2}}{2\pi^{2}\hbar^{2}}\mathbf{b}\cdot\mathbf{B}$ & \cite{zyuzin2012prb}\tabularnewline
\hline
chiral magnetic effect & $\mathbf{j}=\frac{ne^{2}}{2\pi^{2}\hbar^{2}}b_{0}\mathbf{B}$ & \cite{fukushima2008prd,vilenkin1980prd}\tabularnewline
\hline
AHE & $\mathbf{j}=-\frac{ne^{2}}{2\pi^{2}\hbar^{2}}\mathbf{b}\times\mathbf{E}$ & \cite{shindou2001prl,yang2011prb}\tabularnewline
\hline
chiral pseudomagnetic effect & $\mathbf{j}=\frac{ne^{2}}{2\pi^{2}\hbar^{2}}m_{0}\mathbf{B}_{5}$ & \cite{zhou2013cpl,Sumiyoshi2016PRL,grushin2016prx,pikulin2016prx}\tabularnewline
\hline
axial charge density & $j^{50}=\frac{ne^{2}}{2\pi^{2}\hbar^{2}}\mathbf{b}\cdot\mathbf{B}_{5}$ & \tabularnewline
\hline
chiral separation effect & $\mathbf{j^{5}}=\frac{ne^{2}}{2\pi^{2}\hbar^{2}}m_{0}\mathbf{B}$ & \cite{son2004prd,metlitski2005prd,fukushima2008prd,zhou2013cpl}\tabularnewline
\hline
axial pseudoseparation effect & $\mathbf{j^{5}}=\frac{ne^{2}}{2\pi^{2}\hbar^{2}}b_{0}\mathbf{B}_{5}$ & this paper\tabularnewline
\hline
 anomalous axial Hall effect & $\mathbf{j^{5}}=-\frac{ne^{2}}{2\pi^{2}\hbar^{2}}\mathbf{b}\times\mathbf{E}_{5}$ & this paper\tabularnewline
\hline
\end{tabular}

\caption{Topological responses in both single- and multi-Weyl semimetals. The
``Form'' column is valid for both single- and multi-Weyl semimetals
in Eq. $\left(\ref{eq:hamiltonian}\right)$; the ``References''
column is for the corresponding references on single-Weyl semimetals.
(We adopt the Gauss unit in this table so as to connect with results
in other references.)}
\label{topologicalresp}
\end{table}

\end{widetext}

\begin{figure}
\includegraphics[scale=0.33]{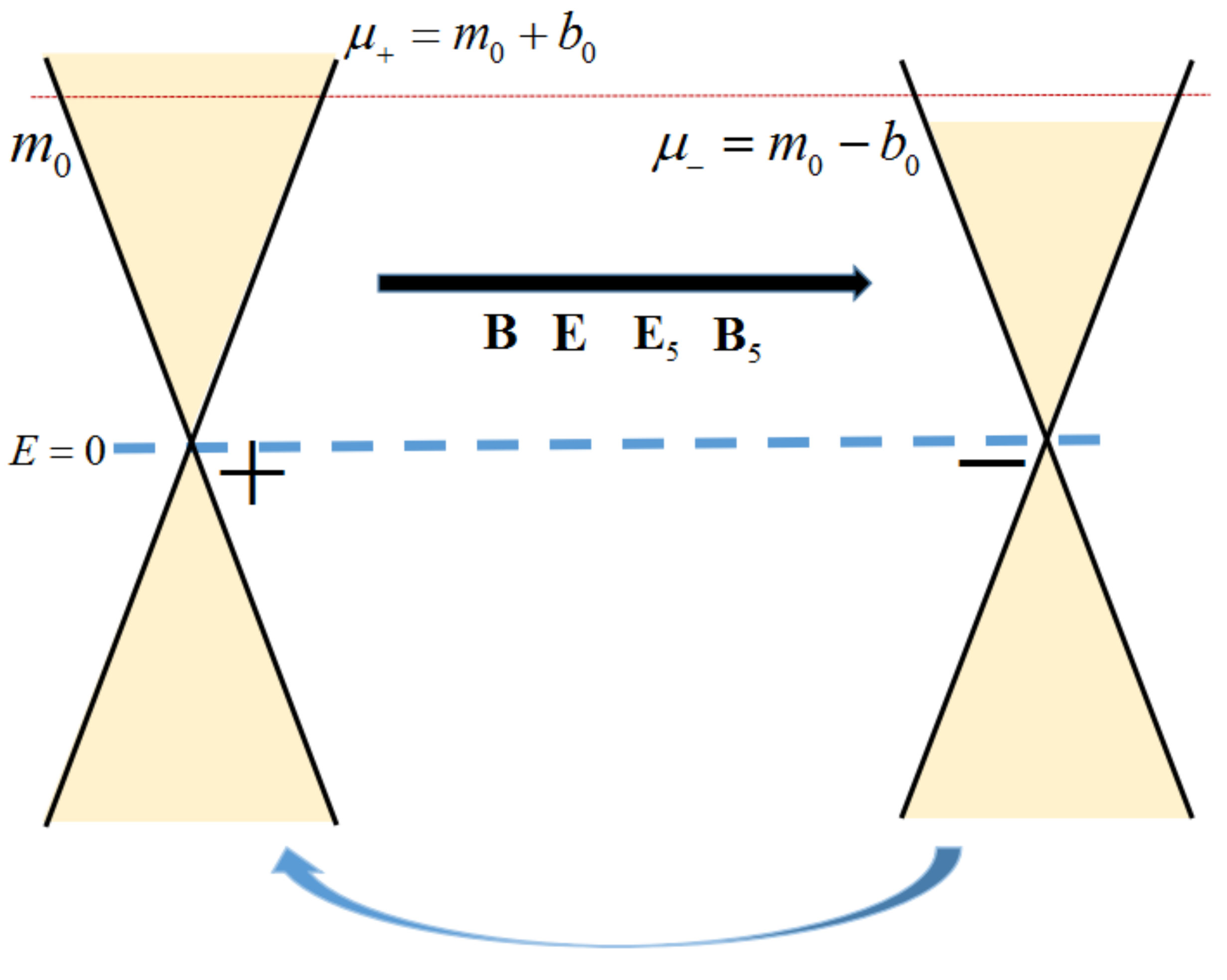}

\caption{Chiral anomaly in the presence of both the external electromagnetic
fields $\mathbf{E}$ and $\mathbf{B}$ as well as the axial fields
$\mathbf{E}_{5}$ and $\mathbf{B}_{5}$. $\pm$ refer to the chirality
of Weyl nodes. The blue arrow denotes for the charge transfer between
two Weyl nodes with opposite chirality. \label{fig:Energy-balance}}
\end{figure}

\section{Chiral anomaly in multi-Weyl semimetals\label{sec:Chiral-anomaly}}

In this section, the equations for the chiral anomaly in multi-Weyl
semimetals in the presence of both electromagnetic fields and axial
fields are derived by using the Fujikawa's method \cite{fujikawa2004oxford}.
For a heuristic purpose, we outline the key points in the application
of the Fujikawa's method in derivation of the chiral anomaly for relativistic
fermions. Since the Jacobian of the chiral-transformation is divergent
like $\ln J\propto\text{Tr}\left[\gamma^{5}\delta(x-x)\right]$ \cite{srednicki2007cambridge},
one needs to introduce a regulator, i.e. $\lim_{M\rightarrow\infty}\exp(-|\slashed{D}|^{2}/M^{2})$
where $\slashed{D}=\gamma^{\mu}D_{\mu}$ is the Dirac operator and
$M$ is a large positive parameter for regularization. After long
and straightforward calculations, one gets $\ln J=-ie^{2}\epsilon^{\mu\nu\alpha\beta}F_{\mu\nu}F_{\alpha\beta}/16\pi^{2}$,
where $\epsilon^{\mu\nu\rho\sigma}$ is the four-rank antisymmetric
Levi-Civita symbol.

Similarly, for multi-Weyl semimetals under electromagnetic fields,
the regulated Jacobian reads
\begin{equation}
J[\beta]=\exp\left\{ -2i\beta\int\lim_{M\rightarrow\infty}\text{tr }\left[\gamma^{5}e^{-\frac{\left|\slashed{iD^{n}}\right|^{2}}{M^{2}}}\delta(x-x)\right]\right\} ,
\end{equation}
where $\beta$ comes from chiral transformation of $\Psi\rightarrow e^{i\beta\gamma^{5}}\Psi$,
and the operator $\slashed{iD^{n}}$ is defined as
\begin{align}
iD_{0,3}^{n} & \left(A\right)=p_{0,3}-eA_{0,3}
\end{align}
and
\begin{align}
iD_{\pm}^{n}\left(A\right) & =\left(p_{\pm}-eA_{\pm}\right)^{n}/w
\end{align}
with $A_{\pm}=\left(A_{1}\pm iA_{2}\right)/\sqrt{2}$ being the components
of the vector potential. By performing a Fourier's transformation
and a Wick's rotation to the Euclidean spacetime, the Jacobian can
be recast as
\begin{eqnarray}
\frac{\ln J[\beta]}{-2i\beta} & = & \epsilon_{E}^{\mu\nu\rho\sigma}\lim_{M\rightarrow\infty}\int\frac{d^{d}k}{(2\pi)^{d}}e^{-\frac{k_{0}^{2}+k_{3}^{2}+\left(k_{1}^{2}+k_{2}^{2}\right)^{n}/2^{n-1}w^{2}}{M^{2}}}\nonumber \\
 &  & \times\frac{\left[iD_{\mu}^{n},\ iD_{\nu}^{n}\right]\left[iD_{\rho}^{n},\ iD_{\sigma}^{n}\right]}{4M^{4}},
\end{eqnarray}
where we have used the identity $4\epsilon_{E}^{\mu\nu\rho\sigma}=\text{tr}\left(\gamma^{5}\gamma_{E}^{\mu}\gamma_{E}^{\nu}\gamma_{E}^{\rho}\gamma_{E}^{\sigma}\right)$,
the subscript $E$ denotes for the Euclidean spacetime. Rescaling
momentum variables as $k_{0,\ 3}\rightarrow Mk_{0,\ 3}$ and $k_{1,\ 2}\rightarrow\left(wM\right)^{1/n}k_{1,\ 2}$,
one finds
\begin{eqnarray}
\frac{\ln J}{-2i\beta} & = & \frac{\epsilon_{E}^{\mu\nu\rho\sigma}}{2}\lim_{M\rightarrow\infty}\int\frac{d^{d}k}{(2\pi)^{d}}e^{-\left[k_{0}^{2}+k_{3}^{2}+\left(k_{1}^{2}+k_{2}^{2}\right)^{n}/2^{n-1}\right]}\nonumber \\
 &  & \times\left[\sum_{m=1}^{n-1}T_{m}\left(wM\right)^{-2m/n}F_{\mu\nu}\left(\partial_{+}\partial_{-}\right)^{m}F_{\rho\sigma}\right.\nonumber \\
 &  & \times\left.\left|D_{+}D_{-}\right|^{n-m-1}-n^{2}F_{\mu\nu}F_{\rho\sigma}\left|D_{+}D_{-}\right|^{n-1}\right],\label{eq:regularization}
\end{eqnarray}
where $T_{m}$ is a combination coefficient, for example, $T_{m}=\delta_{m1}$
for double-Weyl semimetals. Several remarks about the manifestations
of Lorentz symmetry breaking in this derivation are in order here.
First, the decay factor of $\left\{ k_{0}^{2}+k_{3}^{2}+2w^{-2}\left[\left(k_{1}^{2}+k_{2}^{2}\right)/2\right]^{n}\right\} /M^{2}$
instead of $k_{\text{\ensuremath{\mu}}}k^{\mu}/M^{2}$ in single-Weyl
semimetals requires $k_{\mu}$ to be scaled anisotropically. Second,
commutators of covariant derivative are not covariant under Lorentz
transformation. For example, $w\left[iD_{0}^{n=2},\ iD_{\pm}^{n=2}\right]=2F_{0\pm}D_{\pm}+\left(\partial_{\pm}F_{0\pm}\right)$
for double-Weyl semimetals rather than $\left[D_{\mu},\thinspace D_{\nu}\right]=ieF_{\mu\nu}$
for single-Weyl semimetals. As a result, there exists a factor of
$\left|D_{+}D_{-}\right|^{n-1}$ in the second term in the third line,
which disappears in single-Weyl semimetals. In addition, the group
of terms in the second line in Eq. $\left(\ref{eq:regularization}\right)$
completely originates from the breaking of Lorentz symmetry and thus
is forbidden in single-Weyl semimetals.

After taking the limit $M\rightarrow\infty$, all the terms in the
second line in Eq. $\left(\ref{eq:regularization}\right)$ are suppressed.
Thus, one only needs to pick up the leading term of order $M^{0}$
and finally gets the Jacobian for multi-Weyl semimetals as (the derivations
are given in Appendixes \ref{sec:Jacobian-for-chiral} and \ref{sec:jacobian})
\begin{eqnarray}
\ln J[\beta] & = & -i\frac{ne^{2}}{16\pi^{2}}\int\beta\epsilon^{\mu\nu\alpha\beta}F_{\mu\nu}F_{\alpha\beta},\label{eq:Jacobian_electro}
\end{eqnarray}
where coefficient $n$ refers to the winding number defined in Eq.
(\ref{eq:windernumber}). Interestingly, Eq. (\ref{eq:Jacobian_electro})
differs from the counterpart of single-Weyl semimetals by a factor
of $n$. It is consistent with the intuitive picture of the chiral
anomaly in the language of the chiral zeroth Landau levels in the
presence of a magnetic field along the $z$ axis. In fact, there are
$n$ chiral Landau levels crossing the zero energy for multi-Weyl
semimetals in Eq. $\left(\ref{eq:hamiltonian}\right)$ \cite{lixiao2016prb}.

Let us turn to evaluate the Jacobian in the presence of both electromagnetic
fields and axial fields. In order to derive the corresponding Jacobian
from the one only in the presence of electromagnetic fields in Eq.
$\left(\ref{eq:Jacobian_electro}\right)$, we consider the following model
\begin{equation}
\mathcal{L}_{\mathrm{I}}=\sum_{a=1,2}\bar{\Psi}_{a}\gamma^{\mu}iD_{\mu}^{n}\left[A+\left(-1\right)^{a}A^{5}\right]\Psi_{a},\label{eq:L1-1}
\end{equation}
where $\Psi_{a}$ denotes for the four-component Dirac spinor $(\Psi_{a+},\thinspace\Psi_{a-})^{T}$,
subscript $a=1,2$ labels these two Dirac spinors, and $\pm$ stands
for valley or Weyl node degree of freedom. Obviously, there are no
axial gauge fields that couple to $\Psi_{1}$ and $\Psi_{2}$, which
implies the $\mathrm{U}(1)$ symmetry. In addition, the effective
electromagnetic fields couple to Dirac spinors $\Psi_{1}$ and $\Psi_{2}$
differently through $A_{\mu}-A_{\mu}^{5}$ and $A_{\mu}+A_{\mu}^{5}$,
respectively. Thus, one could obtain the Jacobians for $\Psi_{1}$
and $\Psi_{2}$ from the one in Eq. $\left(\ref{eq:Jacobian_electro}\right)$.

Alternatively, this Lagrangian density can also be written as follows:
\begin{equation}
\mathcal{L}_{\mathrm{I}\mathrm{I}}=\sum_{a\neq b=1}^{2}\bar{\Psi}_{ab}\gamma^{\mu}iD_{\mu}^{n}\left(A-\epsilon^{ab}A^{5}\gamma^{5}\right)\Psi_{ab},\label{eq:L2}
\end{equation}
where $\Psi_{ab}=\left(\Psi_{a+},\thinspace\Psi_{b-}\right)^{T}$
with $a,b=1,2$ and $a\neq b$. $D_{\mu}^{n}(A\pm A^{5}\gamma^{5})$
is obtained from $D_{\mu}^{n}(A)$ by replacing $A_{\mu}$ with $A_{\mu}\pm A_{\mu}^{5}\gamma^{5}$.
$\epsilon^{12}=-\epsilon^{21}=1$. One can clearly recognize an axial
gauge field $A^{5}\gamma^{5}$. Hence, the Jacobians for fields $\Psi_{ab}$
in $\mathcal{L}_{\mathrm{I}\mathrm{I}}$ under $\mathrm{U}(1)$ and
the chiral transformation are defined as $J_{\mathrm{U}}\left(A-\epsilon^{ab}A^{5}\gamma^{5}\right)$
and $J_{\mathrm{c}}\left(A-\epsilon^{ab}A_{5}\gamma^{5}\right)$,
respectively. Since $\mathcal{L}_{\mathrm{I}}$ is equivalent to $\mathcal{L}_{\mathrm{I}\mathrm{I}}$,
they should have the same the variations of action $\delta S_{\mathrm{I}}=\delta S_{\mathrm{I}\mathrm{I}}$
under some transformation. To be specific, under the transformations
of $\Psi_{1}\rightarrow e^{i\beta}\Psi_{1}$ and $\Psi_{2}\rightarrow e^{-i\beta}\Psi_{2}$,
the variations of action $S_{\mathrm{I}}$ and $S_{\mathrm{I}\mathrm{I}}$
are
\begin{eqnarray}
\delta S_{\mathrm{I}}^{\left(1\right)} & = & 0
\end{eqnarray}
and
\begin{eqnarray}
\delta S_{\mathrm{I}\mathrm{I}}^{\left(1\right)} & = & \ln J_{\mathrm{c}}(A-A^{5}\gamma^{5})-\ln J_{\mathrm{c}}(A+A^{5}\gamma^{5}).
\end{eqnarray}
Note that $\delta S_{\mathrm{I}}^{\left(1\right)}=0$ is due to the
$\mathrm{U}(1)$ symmetry in $\mathcal{L}_{\mathrm{I}}$. The implementation
of $\Psi_{1,\thinspace2}\rightarrow e^{i\beta\gamma^{5}}\Psi_{1,\thinspace2}$
leads to
\begin{eqnarray}
\delta S_{\mathrm{I}}^{\left(2\right)} & = & -i\frac{ne^{2}}{8\pi^{2}}\int\beta\epsilon^{\mu\nu\rho\sigma}(F_{\mu\nu}F_{\rho\sigma}+F_{\mu\nu}^{5}F_{\rho\sigma}^{5})
\end{eqnarray}
and
\begin{eqnarray}
\delta S_{\mathrm{I}\mathrm{I}}^{\left(2\right)} & = & \ln J_{\mathrm{c}}(A-A^{5}\gamma^{5})+\ln J_{\mathrm{c}}(A+A^{5}\gamma^{5}).
\end{eqnarray}
Thus, one gets the Jacobian
\begin{equation}
\ln J_{\mathrm{c}}(A\pm A^{5}\gamma^{5})=-\frac{ine^{2}\epsilon^{\mu\nu\rho\sigma}}{16\pi^{2}}\int\beta(F_{\mu\nu}F_{\rho\sigma}+F_{\mu\nu}^{5}F_{\rho\sigma}^{5}).\label{eq:jaco_chiral}
\end{equation}
Similarly, we at first perform transformations $\Psi_{1,2}\rightarrow e^{i\beta}\Psi_{1,2}$
and obtain
\begin{equation}
\delta S=\ln J_{\mathrm{U}}\left(A+A^{5}\gamma^{5}\right)+\ln J_{\mathrm{U}}\left(A-A^{5}\gamma^{5}\right)=0.\label{delSa}
\end{equation}
Then, we carry out another transformations: $\Psi_{1}\rightarrow e^{i\beta\gamma^{5}}\Psi_{1}$
and $\Psi_{2}\rightarrow e^{-i\beta\gamma^{5}}\Psi_{2}$, yielding
\begin{align}
\delta S= & -i\frac{ne^{2}}{16\pi^{2}}\int\beta\epsilon^{\mu\nu\alpha\beta}\left[\left(F_{\mu\nu}-F_{\mu\nu}^{5}\right)\left(F_{\alpha\beta}-F_{\alpha\beta}^{5}\right)\right.\nonumber \\
 & \left.-\left(F_{\mu\nu}+F_{\mu\nu}^{5}\right)\left(F_{\alpha\beta}+F_{\alpha\beta}^{5}\right)\right]\nonumber \\
= & \ln J_{\mathrm{U}}\left(A-A^{5}\gamma^{5}\right)-\ln J_{\mathrm{U}}\left(A+A^{5}\gamma^{5}\right).\label{delSb}
\end{align}
Combining Eq. $\left(\ref{delSa}\right)$ and Eq. $\left(\ref{delSb}\right)$,
one finds the Jacobian for $U\left(1\right)$ transformation
\begin{equation}
\ln J_{\mathrm{U}}(A\mp A^{5}\gamma^{5})=\pm i\frac{ne^{2}\epsilon^{\mu\nu\rho\sigma}}{8\pi^{2}}\int\beta F_{\mu\nu}F_{\rho\sigma}^{5}.
\end{equation}
With the help of $J_{\mathrm{U}}$ and $J_{\mathrm{c}}$, it is straightforward
to derive the continuity equations for the electric current and the
axial current for multi-Weyl semimetals
\begin{eqnarray}
\partial_{\mu}j^{\mu} & = & \frac{ne^{3}}{8\pi^{2}}\epsilon^{\mu\nu\rho\sigma}F_{\mu\nu}F_{\rho\sigma}^{5}\label{eq:anomaly}
\end{eqnarray}
and
\begin{eqnarray}
\partial_{\mu}j^{5\mu} & = & \frac{ne^{3}}{16\pi^{2}}\epsilon^{\mu\nu\rho\sigma}\left(F_{\mu\nu}F_{\rho\sigma}+F_{\mu\nu}^{5}F_{\rho\sigma}^{5}\right),\label{eq:chialanomaly}
\end{eqnarray}
or in terms of $j_{\pm}^{\mu}$ and $\mathbf{E}$, $\mathbf{E}_{5}$,
$\mathbf{B}$, $\mathbf{B}_{5}$,
\begin{align}
\partial_{\mu}j_{s}^{\mu} & =-s\frac{ne^{3}}{4\pi^{2}}\mathbf{E}_{s}\cdot\mathbf{B}_{s},\label{eq:+-}
\end{align}
where $\text{\ensuremath{\mathbf{E}}}_{s}\equiv\mathbf{E}+s\mathbf{E}_{5}$
and $\mathbf{B}_{s}\equiv\mathbf{B}+s\mathbf{B}_{5}$ are the effective
electric field and the effective magnetic field near the Weyl node
$s$. $j_{s}^{\mu}=\left(j_{s}^{0},\mathbf{j}_{s}\right)$ refers
to the current near the Weyl node. Eq.\,\,(\ref{eq:anomaly}) indicates
the breaking of $\mathrm{U}(1)$ symmetry and the local charge non-conservation.
Such an anomalous effect from $\mathbf{E}_{5}\cdot\mathbf{B}$ was
attributed to the modulation of the band structure. To be specific,
the band structure of the bulk is modified by strain fields through
compressing or stretching an infinite crystal such that the chemical
potential varies in order to accommodate these fixed number of electrons
\cite{pikulin2016prx}. The local charge non-conservation due to $\mathbf{E}\cdot\mathbf{B}_{5}$
might originate from the charge transfer between bulk and boundary
\cite{pikulin2016prx}. For $F_{\mu\nu}^{5}=0$, the right hand side
of Eq. $\left(\ref{eq:anomaly}\right)$ vanishes, whereas Eq. $\left(\ref{eq:chialanomaly}\right)$
will reduce to the conventional chiral anomaly equation \cite{bertlmann2000book}.
It is worth noting that this set of anomaly equations is known as
the covariant anomaly in high-energy physics \cite{Bardeen1969wardI,fujikawa2004oxford}
and acts as the starting point to discuss topological responses of
single-Weyl semimetals \cite{liu2013prb,grushin2016prx,pikulin2016prx,Gorbar2017prl}.
It is clear that the chiral anomaly equation in Eq. $\left(\ref{eq:+-}\right)$
differs from the counterparts for single-Weyl semimetals by a winding
number $n$ and paves the way to study the physics induced by the
chiral anomaly in multi-Weyl semimetals. It is one of the main results
in this work.

Actually these anomaly equations in Eq. (\ref{eq:+-}) can be intuitively
understood through the lowest chiral Landau levels \cite{nielsen1983plb}.
For simplicity, we consider the case that the strengths of the electric/magnetic
fields are stronger than those of the axial electric/magnetic fields.
When an effective magnetic field $B_{s}$ is applied along $z$-direction,
the energy dispersion of the $n$-fold degenerate lowest chiral Landau
levels for Weyl node $s$ is $sk_{z}$. According to the semiclassical
equation of motion of electrons, adiabatically turning on an additional
electric field $E_{s}$ along $z$-axis leads to a change of the momentum
of quasiparticles $\triangle p_{z}=-seE_{s}\triangle t$ over a period
of time $\triangle t$. Hence, the total variation of charge density
for all of the $n$ chiral Landau levels is given as
\begin{align}
\frac{\triangle j_{s}^{0}}{\triangle t} & =\left(-e\right)n\left(\frac{\triangle p_{z}}{2\pi\triangle t}\right)\left(-\frac{eB_{s}}{2\pi}\right)\nonumber \\
 & =-s\frac{ne^{3}}{4\pi^{2}}E_{s}B_{s},
\end{align}
where the last set of parentheses in the first line denotes for the
degeneracy of each chiral Landau level. This can be recast in a covariant
form as: $\partial_{\mu}j_{s}^{\mu}=-s\frac{ne^{3}}{4\pi^{2}}\mathbf{E}_{s}\cdot\mathbf{B}_{s}$,
which is identical to Eq. (\ref{eq:+-}). It has also been numerically
demonstrated that even tilting the magnetic field away from $z$-axis,
one still obtains $n$ chiral Landau levels crossing the zero energy
\cite{lixiao2016prb}. Carrying out the similar procedure above, one
finally reproduces Eq. (\ref{eq:+-}) as well. It should be noted
that the chiral anomaly equation in Eq. $\left(\ref{eq:+-}\right)$
can also be straightforwardly verified within the semiclassical chiral
kinetic theory \cite{Xiao10RMP,Son2013PRD}.

\section{topological responses \label{sec:top-resp}}

In this section, the topological responses of multi-Weyl fermions
to the electromagnetic field and the axial electromagnetic field are
derived. In the end, the realizations of the strain-induced AHE and
the anomalous axial Hall effect are proposed in Weyl semimetals.

As a minimal coupling between gauge potentials ($A_{\mu}$ and $A_{\mu}^{5}$)
and currents is also valid for non-relativistic fermions \cite{frohlich1995cmp},
the interaction terms are: $A_{\mu}j^{\mu}$ and $A_{\mu}^{5}j^{5\mu}$.
Therefore, $m_{\mu}j^{\mu}$ and $b_{\mu}j^{5\mu}$ can be replaced
by using Eq.\,\,(\ref{eq:anomaly}) and Eq.\,\,(\ref{eq:chialanomaly}).
For example, $-\int b_{\mu}j^{\mu}=\int\left(b_{\mu}x^{\mu}\right)\partial_{\mu}j^{\mu}$
and the corresponding action is
\begin{align}
S_{\text{eff}} & =-i\frac{ne^{2}}{4\pi^{2}}\epsilon^{\mu\nu\rho\sigma}\int\nonumber \\
 & \times\left[b_{\mu}\left(A_{\nu}\partial_{\rho}A_{\sigma}+A_{\nu}^{5}\partial_{\rho}A_{\sigma}^{5}\right)+2m_{\mu}A_{\nu}\partial_{\rho}A_{\sigma}^{5}\right],\label{eq:effaction}
\end{align}
which captures the topological responses associated with $b_{\mu}$
and $m_{\mu}$. For $n=1$, the second term in Eq. $\left(\ref{eq:effaction}\right)$
is nothing but the phonon Hall viscosity proposed in single-Weyl semimetals
\cite{cortijo2015prl}. The topological responses to the fields can
be obtained by varying the effective action $S_{\text{eff}}$ with
respect to $A_{\mu}$ and $A_{\mu}^{5}$
\begin{eqnarray}
j^{\mu} & = & \frac{ne^{2}}{2\pi^{2}}\epsilon^{\mu\nu\rho\sigma}\left(b_{\nu}\partial_{\rho}A_{\sigma}+m_{\nu}\partial_{\rho}A_{\sigma}^{5}\right)\label{eq:jcovariant}
\end{eqnarray}
and
\begin{eqnarray}
j^{5\mu} & = & \frac{ne^{2}}{2\pi^{2}}\epsilon^{\mu\nu\rho\sigma}\left(b_{\nu}\partial_{\rho}A_{\sigma}^{5}+m_{\nu}\partial_{\rho}A_{\sigma}\right).\label{eq:chiralu1current}
\end{eqnarray}
Note that their explicit expressions in terms of $\mathbf{E}$, $\mathbf{B}$,
$\mathbf{E}_{5}$ and $\mathbf{B}_{5}$ are listed in Table\,\,\ref{topologicalresp}.
It is clear that the axial currents in Eq. (\ref{eq:chiralu1current})
can be obtained from $j^{\mu}$ by interchanging $A_{\sigma}$ with
$A_{\sigma}^{5}$, and vice versa. Alternatively, the current in Eq.
(\ref{eq:jcovariant}) can be written as a sum of a polarization current
and a magnetization current (the details are given in appendix \ref{sec:polarization})
\begin{equation}
\mathbf{j}=\partial_{t}\mathbf{P}+\nabla\times\mathbf{M},\label{eq:polarization_magnetization}
\end{equation}
where the polarization vector $\mathbf{P}$ and the magnetization
vector $\mathbf{M}$ are defined by
\begin{align}
\mathbf{P} & =\frac{ne^{2}}{2\pi^{2}}\left[\left(b_{\alpha}x^{\alpha}\right)\mathbf{B}+\left(m_{\alpha}x^{\alpha}\right)\mathbf{B}_{5}\right]
\end{align}
and
\begin{align}
\mathbf{M} & =\frac{ne^{2}}{2\pi^{2}}\left[\left(b_{\alpha}x^{\alpha}\right)\mathbf{E}+\left(m_{\alpha}x^{\alpha}\right)\mathbf{E}_{5}\right],
\end{align}
respectively. For $n=1$, setting the axion field $b_{\alpha}x^{\alpha}=-\pi/2$
and the axial electromagnetic fields $\mathbf{B}_{5}=0$ and $\mathbf{E}_{5}=0$,
one yields $\mathbf{P}=-e^{2}\mathbf{B}/4\pi$, which is exactly the
topological magnetoelectric effect \cite{QiXL2008prb}. Specifically,
the charge density and the current density can be recast as
\begin{eqnarray}
j^{0} & = & \frac{ne^{2}}{2\pi^{2}}\left(\mathbf{b}\cdot\mathbf{B}+\mathbf{m}\cdot\mathbf{B}_{5}\right)\label{rhoden}
\end{eqnarray}
and
\begin{eqnarray}
\mathbf{j} & = & \frac{ne^{2}}{2\pi^{2}}\left(b_{0}\mathbf{B}+m_{0}\mathbf{B}_{5}-\mathbf{b}\times\mathbf{E}-\mathbf{m}\times\mathbf{E}_{5}\right).\label{jden}
\end{eqnarray}
The axial charge and current densities are given as
\begin{eqnarray}
j^{50} & = & \frac{ne^{2}}{2\pi^{2}}\left(\mathbf{b}\cdot\mathbf{B}_{5}+\mathbf{m}\cdot\mathbf{B}\right)\label{rho5den}
\end{eqnarray}
and
\begin{eqnarray}
\mathbf{j^{5}} & = & \frac{ne^{2}}{2\pi^{2}}\left(m_{0}\mathbf{B}+b_{0}\mathbf{B}_{5}-\mathbf{m}\times\mathbf{E}-\mathbf{b}\times\mathbf{E}_{5}\right).\label{j5den}
\end{eqnarray}
The first term in Eq. (\ref{jden}) is the celebrated chiral magnetic
effect \cite{vilenkin1980prd}. The second term relates to the newly-predicted
chiral pseudomagnetic effect \cite{zhou2013cpl,Sumiyoshi2016PRL,grushin2016prx,pikulin2016prx},
which can enhance the magnetoconductivity \cite{grushin2016prx}.
The third term in Eq. (\ref{jden}) is the AHE \cite{Nagaosa2010RMP}.
It is known, the two-dimensional integer Hall conductance (with a
unit normal vector $\hat{\mathbf{k}}$) can be written as: $\mathbf{j}=-\frac{ne^{2}}{2\pi^{2}}\hat{\mathbf{k}}\times\mathbf{E}$.
Therefore, from the viewpoint of the anomalous Hall conductivity,
$\mathbf{j}=-\frac{ne^{2}}{2\pi^{2}}\mathbf{b}\times\mathbf{E}$ indicates
the possible connection between the Weyl semimetals without time reversal
symmetry and the two-dimensional Chern insulators.

The first term in Eq. (\ref{j5den}) is known as the chiral separation
effect in quantum chromodynamics \cite{son2004prd,metlitski2005prd}
and the analogue of the valley current in valleytronics \cite{zhou2013cpl}.
It can also be understood as follows: under an external magnetic field,
the right-handed fermions and the left-handed fermions move parallel
to and antiparallel to the magnetic field, due to opposite chirality
\cite{nielsen1983plb,zhou2013cpl}. To the best of our knowledge,
the second term is firstly derived in this paper and needs a nonzero
chiral chemical potential $b_{0}$. Physically, the axial magnetic
field $\mathbf{B}_{5}$ initially induces an electric current for
both left-handed fermions and right-handed fermions \cite{zhou2013cpl}.
An extra negative sign from $b_{0}$ leads to an axial current. It
is our second main result. The final term in Eq. (\ref{j5den}) can
be regarded as a cousin of AHE. It originates from the fact that $\mathbf{E}_{5}$
couples with opposite signs to the left- and right-handed Weyl fermions.
The occurrence of all the terms linear in $\mathbf{m}$ above requires
the nonzero sum of center of momentum of all Weyl nodes.

All the terms in Eq. (\ref{eq:jcovariant}) and Eq. (\ref{eq:chiralu1current})
can also be obtained in a physically intuitive way, i.e. the energy
balance argument \cite{nielsen1983plb} and the force balance argument.
We first focus on physics near Weyl node $+$ with an effective chemical
potential $m_{0}+b_{0}$. When both electromagnetic and axial electromagnetic
fields are turned on, Eq. (\ref{eq:+-}) seemingly implies that there
are quasiparticles with energy $m_{0}+b_{0}$ created from the Dirac
sea. On the other hand, the quasiparticles with an opposite chirality
annihilate (see Fig. (\ref{fig:Energy-balance})). Therefore, we generalize
the elegant energy balance argument \cite{nielsen1983plb} to the
present case with both electromagnetic fields and axial fields
\begin{align}
 & \mathbf{j}_{+}\cdot\mathbf{E}_{+}+\mathbf{j}_{-}\cdot\mathbf{E}_{-}\nonumber \\
= & \frac{ne^{2}}{4\pi^{2}}\left[\left(m_{0}+b_{0}\right)\mathbf{E}_{+}\cdot\mathbf{B}_{+}-\left(m_{0}-b_{0}\right)\mathbf{E}_{-}\cdot\mathbf{B}_{-}\right].\label{eq:joule}
\end{align}
Physically, the first line is the energy extracted from external fields
and the second line equals the energy due to charge pumping. Note
that the left-hand side of Eq. (\ref{eq:joule}) equals $\mathbf{j}\cdot\mathbf{E}+\mathbf{j}^{5}\cdot\mathbf{E}_{5}$
with the total current $\mathbf{j}=\mathbf{j}_{+}+\mathbf{j}_{-}$
and the axial current $\mathbf{j}^{5}=\mathbf{j}_{+}-\mathbf{j}_{-}$,
which implies that the axial current couples to the axial electric
field and costs energy.

Phenomenologically, because of the non-zero averaged momentum of quasiparticles
near Weyl nodes $\pm$, energy transfer in Eq. (\ref{eq:joule}) must
accompany with transfer of momentum of quasiparticles (see Fig. \ref{fig:Energy-balance}).
The total forces exerting on quasiparticles should vanish, including
the forces due to momentum transfer, the electric forces and the Lorentz
forces, that is, the force balance condition
\begin{equation}
\sum_{s=\pm1}\left(j_{s}^{0}\text{\ensuremath{\mathbf{E}}}_{s}+\mathbf{j}_{s}\times\mathbf{B}_{s}\right)=\frac{ne^{2}}{4\pi^{2}}\sum_{s=\pm1}s\left(\mathbf{m}+s\mathbf{b}\right)\mathbf{E}_{s}\cdot\mathbf{B}_{s}.\label{eq:force_balance}
\end{equation}
Since both the energy balance argument in Eq. (\ref{eq:joule}) and
the force balance condition in Eq. (\ref{eq:force_balance}) are valid
for arbitrary external fields, one can employ vector analysis to obtain
the corresponding charge density and the current density, which exactly
coincide with those in Eqs. (\ref{rhoden}), (\ref{jden}), (\ref{rho5den})
and (\ref{j5den}).

Note that Eq.\,\,(\ref{eq:jcovariant}) does not obey the continuity
equations in Eqs.\,\,(\ref{eq:anomaly}) and (\ref{eq:chialanomaly}).
To restore the continuity equations above, other terms are needed
to be included. Since $A_{\nu}^{5}$ is observable but $A_{\mu}$
not \cite{cortijo2015prl}, we write these extra terms as
\begin{equation}
\delta j^{\mu}=\frac{ne^{3}}{2\pi^{2}}\epsilon^{\mu\nu\rho\sigma}A_{\nu}^{5}\partial_{\rho}A_{\sigma},\label{CS4C}
\end{equation}
which refers to a Chern-Simons contribution to the current density
in Eq. (\ref{jden}) \cite{Bardeen1969wardI,bardeen1984npb}. Its
spatial components can be recast as
\begin{equation}
\delta\mathbf{j}=\frac{ne^{3}}{2\pi^{2}}\left(A_{0}^{5}\mathbf{B}-\mathbf{A}^{5}\times\mathbf{E}\right),\label{CSC}
\end{equation}
while the temporal component is given as $j^{0}=\frac{ne^{3}}{2\pi^{2}}\mathbf{A}^{5}\cdot\mathbf{B}$.
The first term in Eq. $\left(\ref{CSC}\right)$ is the strain-induced
chiral magnetic effect \cite{cortijo2016prb}, whereas the second
term can be dubbed as the strain-induced AHE. This strain-induced
AHE can be understood by the following argument: the strain fields
achieved from stretching or compressing the sample would alter the
crystal constants in some direction. Consequently, the distance between
the Weyl nodes with opposite chirality in momentum space is effectively
changed by $\mathbf{A}^{5}$, leading to a modification to the anomalous
Hall current. The topological Chern-Simons terms in Eq. $\left(\ref{CS4C}\right)$
are regarded as a ground-state current coming from the carriers far
from the Fermi surface, which are not well described by the effective
Hamiltonian near each Weyl node \cite{liu2013prb,landsteiner2014PRB}.
This scenario has recently been used to construct the semiclassical
chiral kinetic theory to investigate plasmons in Weyl materials \cite{Gorbar2017prl}.
It should be noted that, in confined systems with boundaries (nanowires
or thin films), the local nonconservation of the electric current
in Eq. $\left(\ref{eq:jcovariant}\right)$ is attributed to a charge
pumping between the bulk and the surface \cite{pikulin2016prx}. Thus,
there is no global charge nonconservation.

Before closing this section, let us discuss the realization of the
strain-induced AHE in double-Weyl semimetals (The details are given
in Appendix \ref{sec:Axial-gauge-field}). For the sake of simplicity,
we assume that $\mathbf{b}$ is along the $z$ direction in this case,
that is, $\mathbf{b}=b_{3}\hat{z}$. A longitudinal sound wave with
frequency $\omega$ along $z$ direction can produce a displacement
field, $\mathbf{u}=u_{0}\sin\left(qz-\omega t\right)\hat{z}$, which
gives rise to $\mathbf{A}^{5}=\left(0,\ 0,\ -\frac{1}{ea}\cot\left(ab_{3}\right)u_{0}q\cos\left(qz-\omega t\right)\right)$
($a$ is the lattice constant) \cite{pikulin2016prx}. In the limit
$ab_{3}\ll1$, one approximates $\cot\left(ab_{3}\right)\simeq1/ab_{3}$
and then gets the anomalous Hall conductivity as
\begin{equation}
\sigma_{yx}=\frac{e^{2}u_{0}q}{\pi^{2}a^{2}b_{3}}\cos\left(qz-\omega t\right).
\end{equation}
The estimation the corresponding coefficients is carried out as follows:
$b_{3}\sim2\pi/\chi a$, $u_{0}\sim10^{-2}a$, $c_{s}\sim2.3\times10^{3}$
m/s (sound speed) and $\lambda_{s}\sim11\times10^{-6}$ m (sound wavelength)
\cite{chang2013am,pikulin2016prx}. The wave vector is thus of order:
$q\sim2\pi/\lambda_{s}$ and the axial vector potential: $\mathbf{A}^{5}\sim0.01\chi/e\lambda_{s}$.
Therefore, the Hall conductance: $\sigma_{yx}\sim\frac{e^{2}\chi}{200\pi^{2}\lambda_{s}}$.
The ratio between this one and the conventional anomalous Hall conductance
is of order: $0.01\chi^{2}a/\lambda_{s}$. Since the lattice constant
is of order $10^{-10}\:$m and $\lambda_{s}$ of order $10^{-5}\:$m,
so this effect is comparable with the conventional one if $\chi$
is of order $10^{3}$ or larger.

\begin{figure}
\includegraphics[scale=0.38]{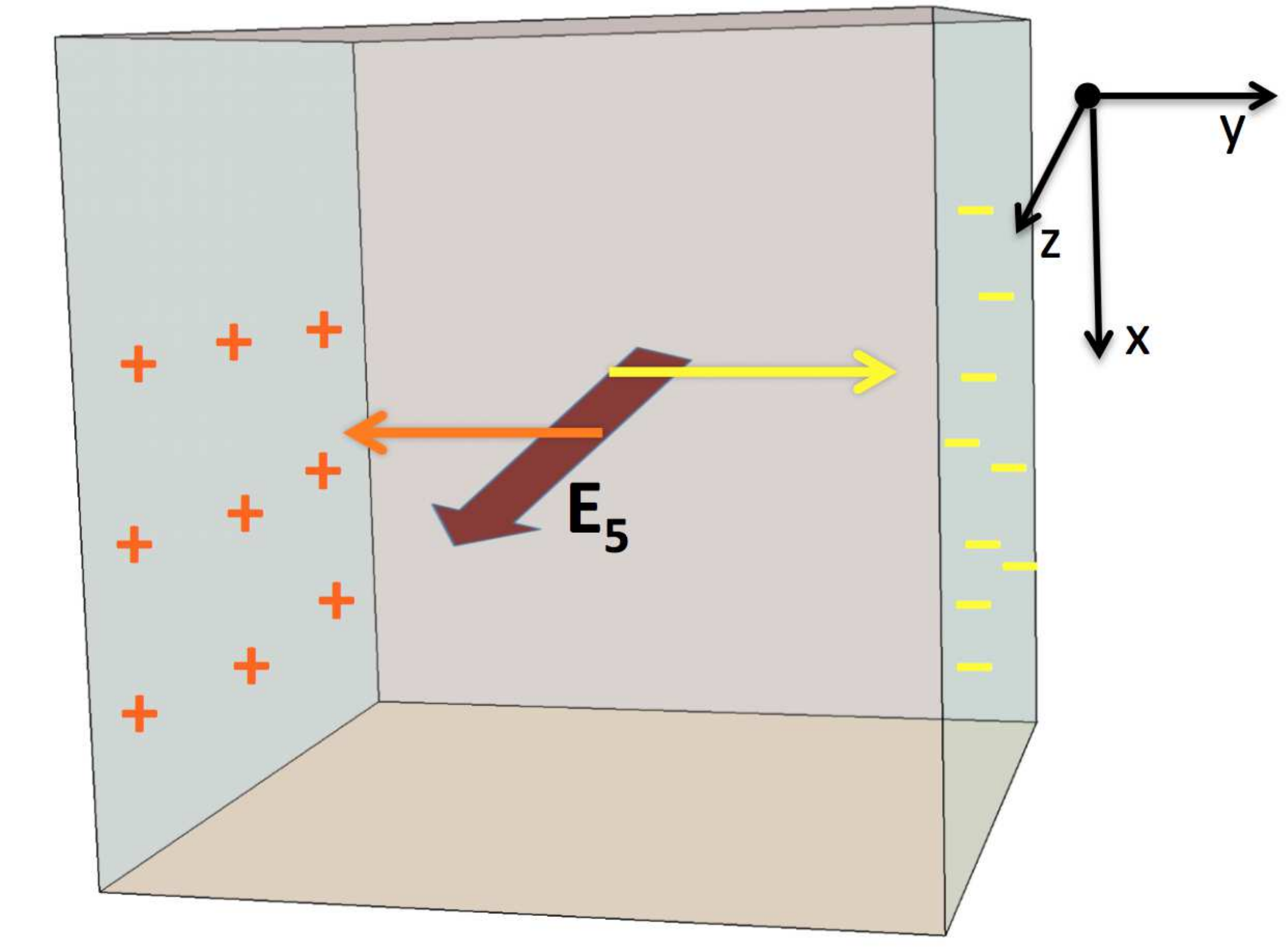}\caption{Schematic figure showing the anomalous axial Hall effect in Table\,\,\ref{topologicalresp}
driven by an axial electric field $\mathbf{E}_{5}$ and the resulting
dynamical chirality accumulation at the surfaces perpendicular to
this axial current. \label{figValleyWeyl}}
\end{figure}

A finite anomalous axial Hall effect in Table\,\,\ref{topologicalresp}
requires an axial electric field that deviates from the direction
of vector $\mathbf{b}=\left(b_{1},\thinspace0,\thinspace b_{3}\right)$.
Following the procedure above, one immediately gets $\mathbf{E}_{5}\left(t\right)=\frac{1}{ea}\cot\left(ab_{3}\right)u_{0}q\omega\sin\left(qz-\omega t\right)\hat{z}$.
Thus, the magnitude of the anomalous axial Hall current in the $y$
direction in the limit $ab_{3}\ll1$ becomes
\begin{equation}
j^{5y}=-\frac{eu_{0}q\omega b_{1}}{\pi^{2}a^{2}b_{3}}\sin\left(qz-\omega t\right),\label{AAHC}
\end{equation}
which leads to the chirality accumulation at the surfaces perpendicular
to the $y$ direction, as shown in Fig. \ref{figValleyWeyl}. It is
our third main result. This effect can be seen as the three-dimensional
counterpart of the valley Hall effect in graphene-like systems \cite{Xiao2007PRL},
in which electrons in different valleys follow in opposite directions
perpendicular to the electric field then accumulate near different
boundaries of systems. It is instructive to compare the chirality
accumulation created by an axial electric field with that induced
by a magnetic field through the chiral separation effect \cite{zhou2013cpl}.
First, the former occurs at the surfaces perpendicular to the cross
product of $\mathbf{b}\times\mathbf{E}_{5}$, while the latter is
at the surfaces perpendicular to the magnetic field. Second, the magnitude
of the one due to the axial electric field is independent of the chemical
potential, whereas that of the magnetic field-induced one is linear
in the chemical potential.

Now we turn to estimate this chirality accumulation by considering
a half-infinity large system locate at $y\geq0$ with open boundary
at $y=0$. Due to the translational symmetry in the $x$-direction,
one thus assumes that the chirality density is independent of $x$.
The continuity equation of the axial current near the surface is modified to
\begin{equation}
\partial_{t}j^{50}\left(y,\thinspace z,\thinspace t\right)=-\frac{1}{\tau_{c}}j^{50}\left(y,\thinspace z,\thinspace t\right)-\nabla\cdot\mathbf{j}^{5}\left(y,\thinspace z,\thinspace t\right),
\end{equation}
where $\tau_{c}$ is the relaxation time of the chirality density
near the surface and $\mathbf{j}^{5}$ comes from the diffusion law
and the anomalous axial Hall current
\begin{equation}
j^{5y}=-D_{5}\partial_{y}j^{50}\left(y,\thinspace z,\thinspace t\right)-\frac{eu_{0}q\omega b_{1}}{\pi^{2}a^{2}b_{3}}\sin\left(qz-\omega t\right),
\end{equation}
where $D_{5}$ is the diffusion parameter. We impose the boundary
condition that $j^{5y}$ vanishes at $y=0$. By combining these two
equations together, one can find
\begin{equation}
\partial_{t}j^{50}\left(y,z,t\right)=-\frac{1}{\tau_{c}}j^{50}\left(y,z,t\right)+D_{5}\partial_{y}^{2}j^{50}\left(y,z,t\right),\label{eq:55}
\end{equation}
which is exactly a diffusion function of the chirality density with
the relaxation time $\tau_{c}$. For simplicity, we consider the limit
of $\omega\ll\tau_{c}^{-1}$ and take following ansatz:
\begin{equation}
j^{50}\left(y,z,t\right)=\text{Im}\left\{ \rho_{a}\left(y\right)\exp\left[i\left(qz-\omega t\right)\right]\right\} .\label{eq:ansatz}
\end{equation}
Note that our ansatz is valid in the limit of $\text{Im }\rho_{a}\left(y\right)\ll\text{Re}\rho_{a}\left(y\right)$.
Inserting Eq. (\ref{eq:ansatz}) into Eq. (\ref{eq:55}) leads to
\begin{equation}
\rho_{a}\left(y\right)=\frac{eu_{0}q\omega b_{1}}{\pi^{2}a^{2}b_{3}\lambda D_{5}}\exp\left(-\lambda y\right),
\end{equation}
where $\lambda\equiv\sqrt{\left(-i\omega+1/\tau_{c}\right)/D_{5}}$
and the factor before $\exp\left(-\lambda y\right)$ is determined
by the boundary conditions: $\rho_{a}\left(y\right)|_{y\rightarrow+\infty}=0$
and $j^{5y}\left(y,\thinspace z,\thinspace t\right)|_{y=0}=0$. In
the limit $\omega\ll\tau_{c}^{-1}$, $j^{50}\left(y,z,t\right)$ can
be further simplified as
\begin{equation}
j^{50}\left(y,z,t\right)=\frac{eu_{0}q\omega b_{1}\sqrt{D_{5}}}{\pi^{2}a^{2}b_{3}\sqrt{\tau}_{c}}e^{\frac{-y}{\sqrt{D_{5}\tau_{c}}}}\sin\left(qz-\omega t\right),
\end{equation}
which suggests that the chirality is confined to a narrow region of
order $\delta y\sim\sqrt{D_{5}\tau_{c}}$.

In reality, there are three characteristic timescales that are associated
with the detection of the chirality accumulation induced by an alternating
axial electric field: period of the axial electric field $2\pi/\omega$,
$\tau_{c}$ and period of the probe light $\tau_{p}$. The period
of the modern ultrafast probe light can be of order femtosecond, while
the frequency of the axial electric field is typically several hundreds
MHz. $\tau_{c}$, a material-dependent parameter, might be dominated
by the internode relaxation time $\tau_{v}$, which has been recently
evaluated from the measurement of magnetotransport in zirconium pentatelluride,
$\tau_{v}^{-1}\sim$ THz \cite{Li2016NP}. The observation of the
chirality accumulation requires these characteristic times satisfy
the condition: $\omega^{-1}\gg\tau_{p}$ and $\tau_{c}\gg\tau_{p}$.
Due to the alternating nature of the axial electric field, there is
no net stationary chirality accumulation during a long time interval.

\section{discussions and conclusions \label{sec:conclusions}}

We perform detailed derivations of the equations of the chiral anomaly
for multi-Weyl semimetals by using the Fujikawa's method and emphasize
several manifestations of the broken Lorentz symmetry due to their
nonlinear nature of energy dispersions. It has also been pointed out
that the equations of the chiral anomaly for multi-Weyl semimetals
differ from those for single-Weyl semimetals by the higher winding
number $n$. It can be understood from the intuitive picture of the
chiral anomaly in the language of the chiral Landau levels.

Compared with single-Weyl semimetals, the transport properties of
multi-Weyl semimetals are modified by higher monopole charges or winding
numbers. The axial vector potential contributes a modification to
the anomalous Hall conductivity. Meanwhile, both the axial electric
and the axial magnetic fields make contributions to the axial current.
These axial currents may produce the chirality accumulation at the
surfaces of a finite-size sample even in the absence of external magnetic
fields. Our nonmagnetic mechanism of the dynamical chirality accumulation
may possess great advantages in the application of Weyl semimetals
in promising valleytronics \cite{Rycerz2007NP,Xiao2007PRL}. On the
other hand, the chirality accumulation in Weyl semimetals induced
by a magnetic field through the chiral separation effect is difficult
to identify unambiguously in circular dichroism spectroscopy. The
impacts of impurities scattering and realistic boundary conditions
of Weyl semimetals on a chirality accumulation are also critical issues
for both the experimental detection and potential applications.

In addition, many topological responses are summarized in Table \ref{topologicalresp},
such as AHE, chiral magnetic effect, chiral pseudomagnetic effect
and chiral separation effect, axial pseudoseparation effect and anomalous
axial Hall effect. Finally, it should be noted that the realizations
of general axial gauge fields in multi-Weyl semimetals depend on the
specific properties of materials under consideration, which deserve
for further study in the future.

The authors would like to thank Xiao Li, Dmitry Pikulin, Min-Fong
Yang and Song-Bo Zhang for helpful discussions and comments. We also
thank Kai-Liang Huang for preparing the figures. This work was supported
by the Research Grant Council, University Grants Committee, Hong Kong
under Grant No. 17301116 and C6026-16W.

\appendix

\section{Jacobian for chiral transformation\label{sec:Jacobian-for-chiral}}

In this part, we provide detailed derivations of the Jacobian under
an infinitesimal chiral transformation. For convenience, we focus
on double-Weyl semimetals and then generalize to multi-Weyl semimetals.
We also set $e=1$ hereafter.

\subsection{Model for multi-Weyl semimetals}

The effective Hamiltonian for multi-Weyl semimetals with a pair of
Weyl nodes reads
\begin{eqnarray}
H_{s} & = & s\left[\mathbf{p}_{3}\sigma^{3}+w^{-1}\left(\mathbf{p}_{+}^{n}\sigma^{-}+\mathbf{p}_{-}^{n}\sigma^{+}\right)\right],
\end{eqnarray}
where $n=2$ is the topological charge, $s=\pm1$ are the chirality
of Weyl nodes, $p_{\pm}=(p_{1}\pm ip_{2})/\sqrt{2}$ and $\sigma^{\pm}=(\sigma^{1}\pm i\sigma^{2})/\sqrt{2}$.
Consequently, we have
\[
\{\sigma^{3},\sigma^{\pm}\}=\{\sigma^{+},\sigma^{+}\}=\{\sigma^{-},\sigma^{-}\}=0
\]
and

\[
\{\sigma^{+},\sigma^{-}\}=\{\sigma^{-},\sigma^{+}\}=\{\sigma^{3},\sigma^{3}\}=2.
\]
The corresponding Lagrangian density is given as
\begin{eqnarray}
\mathcal{L} & = & \Sigma_{s}\Psi_{s}^{\dagger}(p_{0}-H_{s})\Psi_{s}\nonumber \\
 & = & \bar{\Psi}\left[p_{0}\gamma^{0}+p_{3}\gamma^{3}+w^{-1}\left(p_{+}^{n}\gamma^{+}+p_{-}^{n}\gamma^{-}\right)\right]\Psi
\end{eqnarray}
with $\Psi=(\Psi_{+},\ \Psi_{-})^{T}$ , $\bar{\Psi}=\Psi^{\dagger}\gamma^{0}$
, $\gamma^{\pm}=\left(\gamma^{1}\mp i\gamma^{2}\right)/\sqrt{2}$
and $(\gamma^{\pm})^{\dagger}=-\gamma^{\mp}$. For later convenience,
we shall turn to the Euclidean spacetime.

\subsection{The Euclidean-spacetime action}

Performing a Wick's rotation, i.e. $p_{0}=i\omega$, one gets the
partition function
\begin{eqnarray}
Z & = & \int\mathcal{D}\left[\bar{\Psi},\Psi\right]\exp\{i\int d^{d}x_{E}\nonumber \\
 &  & \times\bar{\Psi}\left[\omega\gamma_{E}^{0}+p_{3}\gamma_{E}^{3}+w^{-1}\left(p_{+}^{n}\gamma_{E}^{+}+p_{-}^{n}\gamma_{E}^{-}\right)\right]\Psi\},
\end{eqnarray}
where $\gamma_{E}^{0}=\gamma^{0}$, $\gamma_{E}^{3}=-i\gamma^{3}$,
$\gamma_{E}^{\pm}=-i\gamma^{\pm}$ , satisfying $\{\gamma_{E}^{\alpha},\ \gamma_{E}^{\beta}\}=2g_{E}^{\alpha\beta}$.
The metric tensor is given as
\begin{equation}
g_{E}^{\alpha\beta}=\left(\begin{array}{cccc}
1 & 0 & 0 & 0\\
0 & 0 & 1 & 0\\
0 & 1 & 0 & 0\\
0 & 0 & 0 & 1
\end{array}\right)
\end{equation}
with $\alpha,\beta=0,+,-,3$. Therefore, $\slashed{iD^{n}}=iD_{\alpha}^{n}\gamma_{E}^{\alpha}$
is a Hermitian operator with components:
\begin{eqnarray}
iD_{0}^{n}\left(A\right) & = & \omega-A_{0},\nonumber \\
iD_{\pm}^{n}\left(A\right) & = & (p_{\pm}-A_{\pm})^{n}/w,\nonumber \\
iD_{3}^{n}\left(A\right) & = & p_{3}-A_{3}.
\end{eqnarray}

\subsection{Fujikawa's method and chiral transformation}

Following the standard and lengthy derivation \cite{srednicki2007cambridge,bertlmann2000book},
one gets the measure for a global chiral transformation $\Psi^{\prime}=\exp(i\beta\gamma^{5})\Psi$ as
\begin{eqnarray}
\mathcal{D}\bar{\Psi}^{\prime}\mathcal{D}\Psi^{\prime} & = & J[\beta]\mathcal{D}\bar{\Psi}\mathcal{D}\Psi,
\end{eqnarray}
where
\begin{align}
J\left[\beta\right]= & \exp\left[-2i\beta\int\lim_{M\rightarrow\infty}\text{tr }\gamma^{5}e^{-\left|\slashed{iD^{n}}\right|^{2}/M^{2}}\delta(x-x)\right]
\end{align}
with
\[
\left|\slashed{iD^{n}}\right|^{2}=g_{E}^{\alpha\beta}(iD_{\alpha}^{n=2})(iD_{\beta}^{n=2})+\frac{1}{2}\gamma_{E}^{\alpha}\gamma_{E}^{\beta}\left[iD_{\alpha}^{n=2},\ iD_{\beta}^{n=2}\right].
\]
Defining $4\epsilon_{E}^{\alpha_{1}\alpha_{2}\alpha_{3}\alpha_{4}}=\text{tr}\left(\gamma^{5}\gamma_{E}^{\alpha_{1}}\gamma_{E}^{\alpha_{2}}\gamma_{E}^{\alpha_{3}}\gamma_{E}^{\alpha_{4}}\right)$
and performing a Fourier transformation
\begin{align}
\lim_{x\rightarrow y}e^{-\left|iD^{n=2}\right|^{2}/M^{2}}\delta(x-y) & =\int\frac{d^{d}k}{(2\pi)^{d}}e^{-\left|iD^{n=2}\right|^{2}/M^{2}}\nonumber \\
 & \left(p\rightarrow p-k\right),
\end{align}
we get the Jacobian as
\begin{eqnarray}
\frac{\ln[J]}{-2i\beta} & = & \lim_{M\rightarrow\infty}\int\frac{d^{d}k}{(2\pi)^{d}}\epsilon_{E}^{\alpha_{1}\alpha_{2}\alpha_{3}\alpha_{4}}e^{-\frac{k_{0}^{2}+k_{3}^{2}+\left(k_{1}^{2}+k_{2}^{2}\right)^{2}/2w^{2}}{M^{2}}}\nonumber \\
 &  & \frac{\left[iD_{\alpha_{1}}^{n=2},\ iD_{\alpha_{2}}^{n=2}\right]\left[iD_{\alpha_{3}}^{n=2},\ iD_{\alpha_{4}}^{n=2}\right]}{4M^{4}}.
\end{eqnarray}
Note that other terms vanish due to the infinitely large $M$ and
the trace of gamma matrices. The factor of $\exp\left\{ -\left[k_{0}^{2}+k_{3}^{2}+\left(k_{1}^{2}+k_{2}^{2}\right)^{2}/2w^{2}\right]/M^{2}\right\} $
comes from $g_{E}^{\alpha\beta}(iD_{\alpha}^{n=2})(iD_{\beta}^{n=2})$.
As for the commutators, they are given as
\begin{widetext}
\begin{eqnarray}
w^{2}[iD_{0}^{n=2},\ iD_{3}^{n=2}][iD_{-}^{n=2},\ iD_{+}^{n=2}] & = & -iF_{03}\left\{ i\partial_{+}\partial_{-}F_{-+}+2\left[\left(\partial_{-}F_{-+}\right)iD_{+}+\left(\partial_{+}F_{-+}\right)iD_{-}\right]+2iF_{-+}\left\{ D_{+},\ D_{-}\right\} \right\} ,\nonumber \\
w^{2}[iD_{-}^{n=2},\ iD_{+}^{n=2}][iD_{0}^{n=2},\ iD_{3}^{n=2}] & = & \left(\partial_{+}\partial_{-}F_{-+}\right)F_{03}+\left[2\left(\partial_{-}F_{-+}\right)\left(\partial_{+}F_{03}\right)+2\left(\partial_{+}F_{-+}\right)\left(\partial_{-}F_{03}\right)\right]\nonumber \\
 &  & +\left[4F_{-+}\left(\partial_{+}\partial_{-}F_{03}\right)+2F_{-+}F_{03}\left\{ D_{+},\ D_{-}\right\} \right],\nonumber \\
w^{2}[iD_{0}^{n=2},\ iD_{+}^{n=2}][iD_{3}^{n=2},\ iD_{-}^{n=2}] & = & (\partial_{+}F_{0+})(\partial_{-}F_{3-})+2F_{0+}\left(\partial_{+}\partial_{-}F_{3-}\right)+4F_{0+}F_{3-}D_{+}D_{-}.\label{eq:commutators}
\end{eqnarray}
Note that one can exchange subscripts $+$ and $-$ to obtain all
the rest terms.

It is clear that the product of commutators contain not only functions,
but also operators, e.g. $\partial F_{\alpha\beta}D_{\pm}$ and $F_{\alpha\beta}F_{\mu\nu}|D_{\pm}|^{2}$.
This is due to the quadratic dispersion of double-Weyl semimetals.
However, we shall show that only terms containing $|D_{\pm}|^{2}$
survive in the limit of $M\rightarrow\infty$. We first focus on the
part contributed by $|D_{\pm}|^{2}$ :
\begin{eqnarray}
\frac{\ln J}{-2i\beta} & = & \frac{1}{2}\left(4\epsilon_{E}^{\alpha_{1}\alpha_{2}\alpha_{3}\alpha_{4}}\right)\lim_{M\rightarrow\infty}\int\frac{d^{d}k}{(2\pi)^{d}}M^{3}we^{-\left[k_{0}^{2}+k_{3}^{2}+\left(k_{1}^{2}+k_{2}^{2}\right)^{2}/2\right]}\nonumber \\
 &  & \times\frac{Mw}{4M^{4}w^{2}}\left[-4F_{\alpha_{1}\alpha_{2}}F_{\alpha_{3}\alpha_{4}}\frac{\left(k_{1}+A_{1}M^{-1/2}w^{-1/2}\right)^{2}+\left(k_{2}+A_{2}M^{-1/2}w^{-1/2}\right)^{2}}{2}\right]\nonumber \\
 & = & \frac{4\epsilon_{E}^{\alpha_{1}\alpha_{2}\alpha_{3}\alpha_{4}}}{8}\lim_{M\rightarrow\infty}\int\frac{d^{d}k}{(2\pi)^{d}}\left[-2F_{\alpha_{1}\alpha_{2}}F_{\alpha_{3}\alpha_{4}}(k_{1}^{2}+k_{2}^{2})\right]e^{-\left[k_{0}^{2}+k_{3}^{2}+\left(k_{1}^{2}+k_{2}^{2}\right)^{2}/2\right]}\nonumber \\
 & = & \frac{1}{16\pi^{2}}\epsilon^{\mu\nu\alpha\beta}F_{\mu\nu}F_{\alpha\beta},
\end{eqnarray}
where we have rescaled $k_{0,\ 3}\rightarrow Mk_{0,\ 3}$ and $k_{1,\ 2}\rightarrow\sqrt{wM}k_{1,\ 2}$.
For $\partial F_{\alpha_{1}\alpha_{2}}\partial F_{\alpha_{3}\alpha_{4}}$,
the integral is proportional to $\frac{1}{Mw}$, and $\frac{1}{\sqrt{Mw}}$
for $F_{\alpha_{1}\alpha_{2}}\partial F_{\alpha_{3}\alpha_{4}}$ .
Therefore, only $F_{\alpha_{1}\alpha_{2}}F_{\alpha_{3}\alpha_{4}}$
survives in the limit $M\rightarrow\infty$. Note that for multi-Weyl
semimetals with winding number $n$, $k_{\mu}$ are scaled as: $k_{0,\ 3}\rightarrow Mk_{0,\ 3}$,
$k_{1,\ 2}\rightarrow M^{1/n}k_{1,\ 2}$ ($w$ is neglected). Thus,
the coefficient for $F_{\alpha_{1}\alpha_{2}}F_{\alpha_{3}\alpha_{4}}$
is proportional to $M^{2+2/n}M^{-4}M^{\frac{2n-2}{n}}=M^{0}$ , where
$M^{2+2/n}$, $M^{-4}$ and $M^{\frac{2n-2}{n}}$ come from rescaling
of $d^{4}k$ , the exponent and $|D_{\pm}|^{2n-2}$, respectively.
All other terms are suppressed by taking the limit of $M\rightarrow\infty$.
Hence, the Jacobian for multi-Weyl semimetals with winding number $n$ is
\begin{eqnarray}
\frac{\ln J}{-2i\beta} & = & \frac{1}{2}\left(4\epsilon_{E}^{\alpha_{1}\alpha_{2}\alpha_{3}\alpha_{4}}\right)\lim_{M\rightarrow\infty}\int\frac{d^{d}k}{(2\pi)^{d}}e^{-\left\{ k_{0}^{2}+k_{3}^{2}+2\left[\left(k_{1}^{2}+k_{2}^{2}\right)/2\right]^{n}\right\} }\nonumber \\
 &  & \times\frac{1}{4}\left\{ -n^{2}F_{\alpha_{1}\alpha_{2}}F_{\alpha_{3}\alpha_{4}}\left[\left(k_{1}^{2}+k_{2}^{2}\right)/2\right]^{n-1}\right\} \nonumber \\
 & = & \frac{n}{32\pi^{2}}\epsilon^{\mu\nu\alpha\beta}F_{\mu\nu}F_{\alpha\beta}.
\end{eqnarray}
Note that the coefficient $n^{2}$ in the second line comes from the
commutator: $[iD_{+}^{n},\thinspace\thinspace iD_{-}^{n}]$ or $[iD_{-}^{n},\thinspace\thinspace iD_{+}^{n}]$.
\end{widetext}

\section{Effective action in the absence of axial gauge fields\label{sec:jacobian}}

Now we shall derive the effective action for topological responses
to electromagnetic fields with two different ways: Fujikawa's method
and a perturbative approach. This enables us to visualize how winding
number manifests itself in the effective action. A chiral transformation
is implemented as follows:
\begin{eqnarray}
\mathcal{L} & = & \bar{\Psi}e^{ib_{\mu}x^{\mu}\gamma^{5}}\{\gamma^{0}(p_{0}+b_{0}\gamma^{5})+\gamma^{3}(p_{3}+b_{3}\gamma^{5})\nonumber \\
 &  & +\frac{1}{w}\left[\gamma^{+}\left(p_{+}+b_{+}\gamma^{5}\right)^{n}+\gamma^{-}\left(p_{-}+b_{-}\gamma^{5}\right)^{n}\right]\}\nonumber \\
 &  & \times e^{ib_{\mu}x^{\mu}\gamma^{5}}\Psi,
\end{eqnarray}
where $b_{\mu}x^{\mu}$ is not an infinitesimal parameter. Therefore,
we iterate a sequence of infinitesimal chiral transformations: $\Psi\rightarrow e^{ib_{\mu}x^{\mu}\gamma^{5}ds}\Psi$,
with $ds$ an infinitesimal parameter. Hence, after a series of infinitesimal
chiral transformations, the Lagrangian density becomes
\begin{eqnarray}
\mathcal{L}\left(s\right) & = & \bar{\Psi}\{\gamma^{0}\left[p_{0}+(1-s)b_{0}\gamma^{5}\right]+\gamma^{3}\left[p_{3}+(1-s)b_{3}\gamma^{5}\right]\nonumber \\
 &  & +\frac{1}{w}\gamma^{+}\left[p_{+}+(1-s)b_{+}\gamma^{5}\right]^{n}\nonumber \\
 &  & +\frac{1}{w}\gamma^{-}\left[p_{-}+(1-s)b_{-}\gamma^{5}\right]^{n}\}\Psi,
\end{eqnarray}
where $b_{\mu}$ is thus eliminated when $s=1$. Then we sum all these
resulting Jacobians up. Because of Eq. (\ref{eq:jaco_chiral}), the
effective action turns out to be
\begin{eqnarray}
S_{\text{eff}} & = & i\frac{n}{16\pi^{2}}\int d^{d}x\int_{0}^{1}ds\left(b_{\mu}x^{\mu}\right)\epsilon^{\mu\nu\rho\sigma}\nonumber \\
 &  & \times\left[F_{\mu\nu}F_{\rho\sigma}+F_{\mu\nu}^{5}\left(s\right)F_{\rho\sigma}^{5}\left(s\right)\right]\nonumber \\
 & = & i\frac{n}{16\pi^{2}}\int d^{d}x\left(b_{\mu}x^{\mu}\right)\epsilon^{\mu\nu\rho\sigma}F_{\mu\nu}F_{\rho\sigma},\label{eq:effective}
\end{eqnarray}
where $F_{\rho\sigma}^{5}\left(s\right)=\left(1-s\right)\left(\partial_{\rho}b_{\sigma}-\partial_{\sigma}b_{\rho}\right)=0$
.

Let us resort to a perturbative approach to emphasize the role of
topology. For simplicity, we set $b_{\mu}=-\delta_{\mu3}b_{3}$. In
the uniform and dc limit of external fields, the coefficient before
$\epsilon^{3\mu\nu\alpha}A_{\mu}\partial_{\nu}A_{\alpha}$ is given by
\begin{align}
\frac{i}{8\pi^{2}}C & =\frac{\epsilon^{3\mu\nu\alpha}}{3!2!}\int\frac{d^{4}q}{(2\pi)^{4}}\times\nonumber \\
 & \text{tr}\left[\left(G\partial_{q_{\mu}}G^{-1}\right)\left(G\partial_{q_{\nu}}G^{-1}\right)\left(G\partial_{q_{\alpha}}G^{-1}\right)\right],
\end{align}
where we have used Ward's identity
\begin{equation}
-i\Gamma^{\mu}(p)=\partial_{p_{\mu}}G^{-1}(p)
\end{equation}
with $\Gamma^{\mu}(p)=\lim_{k\rightarrow0}\Gamma^{\mu}(p+k,\ p)$.
$\Gamma^{\mu}$ and $G(p)$ are the interacting vertex and real-time
fermion Green's function, respectively. Note that this result is actually
analogue to the Chern-Simons term in odd dimensional spacetime \cite{golterman1993plb,niemi1983prl},
for example, $S_{\text{CS}}^{2+1}\propto\left[\int\left(GdG^{-1}\right)^{3}\right]\int\epsilon^{\mu\nu\rho}A_{\mu}\partial_{\nu}A_{\rho}$
in 2+1 dimension.

In addition, $C$ has a close relation with $N(\mathbf{p}_{3})$ in
Eq. $\left(\ref{eq:Chern}\right)$
\begin{equation}
C=\int dp_{3}N(\mathbf{p}_{3})=-2nb,
\end{equation}
where $n$ originating from $N\left(\mathbf{p}_{3}\right)$ is the
winding number. Hence, $n$ in Eq. $\left(\ref{eq:effective}\right)$
is of topological nature.

\section{Realization of axial gauge fields in double-Weyl semimetals\label{sec:Axial-gauge-field}}

In this section, we shall construct the axial gauge fields in double-Weyl
semimetals, which enable us to realize both strain-induced anomalous
Hall effect and anomalous axial Hall effect in double-Weyl semimetals.
We start with following tight-binding Hamiltonian
\begin{eqnarray}
H & = & \left[t_{1}\cos\left(ak_{z}\right)-\triangle\right]\sigma^{3}\nonumber \\
 &  & +t\left\{ \left[\sin\left(ak_{x}\right)-\alpha\sin\left(ak_{z}\right)\right]^{2}-\cos\left(ak_{y}\right)+1\right\} \sigma^{1}\nonumber \\
 &  & +2t\left[\sin\left(ak_{x}\right)-\alpha\sin\left(ak_{z}\right)\right]\sin\left(ak_{y}\right)\sigma^{2},
\end{eqnarray}
where $a$ is the lattice constant, Pauli matrices $\sigma^{i}$ with
$i=1,2,3$ have the same meanings as those in the main text, and a
pair of Weyl nodes locates at $\pm\mathbf{b}$ in the three-dimensional
Brillouin zone. The vector $\pm\mathbf{b}$ is given as
\begin{equation}
\mathbf{b}=\left(b_{1},\thinspace0,\thinspace b_{3}\right),
\end{equation}
with $b_{1}=\alpha\sin\left(ab_{3}\right)$ and $b_{3}=a^{-1}\left|\arccos\left(\frac{\triangle}{t_{1}}\right)\right|$.
The effective velocity along the $z$ direction is $v_{z}=sat_{1}\sin\left(ab_{3}\right)$.

Now we apply a sound wave along the $z$ direction and get a displacement
field
\begin{equation}
\mathbf{u}=u_{0}\sin\left(qz-\omega t\right)\hat{z}.
\end{equation}
Consequently, the hopping constant along the $z$ direction is modified
due to strain fields \cite{cortijo2015prl,shapourian2015prb},
\begin{equation}
t_{1}\sigma_{z}\rightarrow t_{1}\left(1-u_{33}\right)\sigma_{z},
\end{equation}
where $u_{33}=\partial_{z}u_{3}$. Hence, the variation of the Hamiltonian is
\begin{eqnarray}
\delta H & = & -t_{1}u_{33}\cos\left(ak_{z}\right)\sigma_{z}\nonumber \\
 & \simeq & -st_{1}a\sin\left(ab_{3}\right)\left[seu_{33}\frac{\cot\left(ab_{3}\right)}{ae}\right],
\end{eqnarray}
where we have set $k_{z}=b_{3}$ in the last line.

Hence, the Hamiltonian in the continuous limit is
\begin{eqnarray}
H_{s} & = & s\left\{ \left(k_{z}+sb_{3}-seA_{z}^{5}\right)\sigma^{3}+\left[\left(k_{x}+sb_{1}\right)^{2}-k_{y}^{2}\right]\sigma^{1}\right.\nonumber \\
 &  & +\left.2\left(k_{x}+sb_{1}\right)k_{y}\sigma^{2}\right\} ,
\end{eqnarray}
where the effective velocity has been abosrted to the momentum and
the fermion field has been redefined: $\Psi=\left(\psi_{+},\thinspace\sigma^{3}\psi_{-}\right)$.
The axial gauge field becomes
\begin{equation}
A_{z}^{5}=-u_{33}\frac{\cot\left(ab_{3}\right)}{ea}\simeq-u_{33}\frac{1}{ea^{2}b_{3}},
\end{equation}
where $\simeq$ denotes for the limit of $ab_{3}\ll1$.

\section{Polarization current and magnetization current \label{sec:polarization}}

The aim of this section is to show the current density in Eq. $\left(\ref{eq:jcovariant}\right)$
can be expressed as a sum of a polarization current and a magnetization
current. Let us first rewrite the current density as
\begin{equation}
j^{\mu}=\frac{ne^{2}}{2\pi^{2}}\epsilon^{\mu\nu\rho\sigma}\partial_{\nu}\left[\left(b_{\alpha}x^{\alpha}\right)\partial_{\rho}A_{\sigma}+\left(m_{\alpha}x^{\alpha}\right)\partial_{\rho}A_{\sigma}^{5}\right].
\end{equation}
Because of the anti-symmetry property of the Levi-Civita symbol, the
term of $\epsilon^{\mu\text{\ensuremath{\nu}}\rho\sigma}\partial_{\nu}\partial_{\rho}A_{\sigma}^{(5)}$
vanishes identically. Then the $i$-th component of the current density becomes
\begin{eqnarray}
j^{i} & = & \frac{ne^{2}}{2\pi^{2}}\left\{ \epsilon^{i0jk}\partial_{t}\left[\left(b_{\alpha}x^{\alpha}\right)\partial_{j}A_{k}+\left(m_{\alpha}x^{\alpha}\right)\partial_{j}A_{k}^{5}\right]\right.\nonumber \\
 &  & +\epsilon^{ij0k}\partial_{j}\left[\left(b_{\alpha}x^{\alpha}\right)\partial_{t}A_{k}+\left(m_{\alpha}x^{\alpha}\right)\partial_{t}A_{k}^{5}\right]\nonumber \\
 &  & +\left.\epsilon^{ijk0}\partial_{j}\left[\left(b_{\alpha}x^{\alpha}\right)\partial_{k}A_{0}+\left(m_{\alpha}x^{\alpha}\right)\partial_{k}A_{0}^{5}\right]\right\} .
\end{eqnarray}
Defining the $i$-th components of vectors $\mathbf{P}$ and $\mathbf{M}$ as
\begin{align}
P^{i} & =-\frac{ne^{2}}{2\pi^{2}}\epsilon^{0ijk}\left[\left(b_{\alpha}x^{\alpha}\right)\partial_{j}A_{k}+\left(m_{\alpha}x^{\alpha}\right)\partial_{j}A_{k}^{5}\right]\\
M^{i} & =\frac{ne^{2}}{2\pi^{2}}\epsilon^{0ijk}\left[\left(b_{\alpha}x^{\alpha}\right)\left(\partial_{t}A_{k}-\partial_{k}A_{0}\right)\right.\nonumber \\
 & +\left.\left(m_{\alpha}x^{\alpha}\right)\left(\partial_{t}A_{k}^{5}-\partial_{k}A_{0}^{5}\right)\right],
\end{align}
one thus finds
\begin{equation}
\mathbf{j}=\partial_{t}\mathbf{P}+\nabla\times\mathbf{M}.
\end{equation}
Note that the polarization vector $\mathbf{P}$ and the magnetization
vector $\mathbf{M}$ can also be written as
\begin{align}
\mathbf{P} & =\frac{ne^{2}}{2\pi^{2}}\left[\left(b_{\alpha}x^{\alpha}\right)\mathbf{B}+\left(m_{\alpha}x^{\alpha}\right)\mathbf{B}_{5}\right]
\end{align}
and
\begin{align}
\mathbf{M} & =\frac{ne^{2}}{2\pi^{2}}\left[\left(b_{\alpha}x^{\alpha}\right)\mathbf{E}+\left(m_{\alpha}x^{\alpha}\right)\mathbf{E}_{5}\right].
\end{align}

\bibliographystyle{apsrev4-1}

\end{document}